\documentclass[12pt,reqno]{amsart}
\usepackage{latexsym}

\newcommand{\C}{{\mathbb C}}

\renewcommand{\phi}{\varphi}

\newtheorem{theo}{{\sc Theorem}}[section]
\newtheorem{cor}[theo]{{\sc Corollary}}

\newtheorem{prop}[theo]{{\sc Proposition}}

\newenvironment{defin}{\medskip\noindent{\it Definition:\/} }{\medskip}

\title[Note on Quantum Unique Ergodicity ] {Note on Quantum Unique Ergodicity}

\author{Steve Zelditch }
\address{Department of Mathematics, Johns Hopkins University, Baltimore, MD
21218, USA} \email{zelditch@math.jhu.edu}

\thanks{Research partially supported by NSF grant
 DMS-0071358 and by the Clay Mathematics Institute }

\date{\today}

\begin{document}

 \maketitle

 The purpose of this note is to record an
 observation about quantum unique ergodicity (QUE) which is relevant to
 the  recent construction of H. Donnelly \cite{D} of quasi-modes on certain non-positively
 curved surfaces,  and to  similar quasi-mode constructions known for many years
 as bouncing ball modes on Bunimovich stadia  \cite{BSS, H, BZ1, BZ2}.  Our new  observation (Proposition
 \ref{PONE})  is the asymptotic vanishing
 of near off-diagonal matrix elements   for eigenfunctions of  QUE
 systems. As a corollary, we find that
 quantum ergodic (QE) systems possessing quasi-modes with singular limits and
 with a
 limited
 number of frequencies cannot be QUE.

 We begin by recalling that QUE  (for Laplacians) concerns the matrix elements $\langle
 A
 \phi_i, \phi_j \rangle$ of  pseudodifferential
 operators relative to an orthonormal basis $\{\phi_j\}$ of eigenfunctions
 $$\Delta \phi_j = \lambda_j^2 \; \phi_j,\;\;\; \langle \phi_j,
 \phi_k \rangle = 0. $$
 of the Laplacian $\Delta$ of a compact Riemannian manifold $(M,
 g)$.  We denote the spectrum of $\Delta$ by $Sp(\Delta)$. By definition, $\Delta$ is QUE
 if \begin{equation}  \langle
 A
 \phi_j, \phi_j \rangle \to \int_{S^*M} \sigma_A d L\end{equation}
 where $dL$ is the (normalized) Liouville measure on the unit (co-)tangent
 bundle. The term `unique' indicates that no subsequence of density zero of
 eigenfunctions need be removed when taking the limit.

 The main result of this note is that all  off-diagonal terms of QUE
 systems tend to zero if the eigenvalue gaps tend to zero. This
 strengthens the conclusion of \cite{Z} that almost all
 off-diagonal terms (with vanishing gaps) tend to zero in general
 QE situations. As will be seen below, it also provides evidence
 that Donnelly's examples are non-QUE and establishes a
 localization statement of Heller-O'Connor \cite{HO}.

 \begin{prop}  \label{PONE} Suppose that $\Delta$ is QUE. Suppose that
 $\{(\lambda_{i_r}, \lambda_{j_r}), \; i_r \not= j_r\}$ is a sequence of pairs
 of eigenvalues of $\sqrt{\Delta}$ such that
 $\lambda_{i_r} - \lambda_{j_r} \to 0$ as $r \to \infty$.  Then $d \Phi_{i_r, j_r} \to 0$.
 \end{prop}

 \begin{proof}  We define the distributions $d \Phi_{i, j} \in
 {\mathcal D}'(S^*M)$ by
 $$\langle Op(a) \phi_i, \phi_j \rangle = \int_{S^*M} a d \Phi_{i,
 u} $$
 where $a \in C^{\infty}(S^*M).$  Let $\{\lambda_i, \lambda_j\}$ be any
 sequence of pairs with the gap $\lambda_i - \lambda_j \to 0$. It is then known
 that any weak* limit $d \nu $ of the sequence $\{d \Phi_{i,j} \}$ is a measure invariant under the geodesic
 flow \cite{Z, D}. The weak limit is defined by the property that
 \begin{equation} \label{QUE}   \langle A^*A \phi_i, \phi_j \rangle  \to \int_{S^*M}
 |\sigma_A|^2 d \nu. \end{equation}
 If the eigenfunctions are real, then $d \nu$ is a real (signed)
 measure.

 Our first observation is that any such weak limit must be a
 constant multiple of Liouville measure $dL$. Indeed, we first
 have:
 \begin{equation}| \langle A^*A \phi_i, \phi_j \rangle | \leq | \langle A^*A \phi_i, \phi_i \rangle
 |^{1/2} \; | \langle A^*A \phi_j, \phi_j \rangle |^{1/2}.
 \end{equation}
 Taking the limit along the sequence of pairs, we obtain
 \begin{equation} |\int_{S^*M}
 |\sigma_A|^2 d \nu| \leq \int_{S^*M} |\sigma_A|^2 d L.
 \end{equation}
 It follows that $d \nu << d L$ (absolutely continuous). But $dL$
 is an ergodic measure, so if $d \nu = f d L$ is an invariant measure  with $f \in L^1 (d L)$, then $f $ is constant.
 Thus,
 \begin{equation} \label{C} d \nu = C dL, \;\;\; \mbox{for some constant}\; C.  \end{equation}

 We now observe  that $C = 0$ if $\phi_i \bot \phi_j$ (i.e. if $i \not= j)$.  This follows if we substitute $A =
 I$ in (\ref{QUE}), use orthogonality and (\ref{C}).

\end{proof}

This result has implications for the possible  `scarring' of {\it
quasi-modes} of QUE systems.  We first recall that a quasi-mode of
order $s $ for $\Delta$  is a sequence $\{\psi_k\}$ of
$L^2$-normalized functions which solves \begin{equation}
\label{QM} ||(\Delta - \mu_k) \psi_k||_{L^2} = O(\mu_k^{-s/2}),
\end{equation}  for a sequence of quasi-eigenvalues $\mu_k$ (see \cite{CdV}
for background).  In particular a quasi-mode of order $0$
satisfies $||(\Delta - \mu_k) \psi_k||_{L^2} = O(1)$. Such
(relatively low-order) quasi-modes can be easily  constructed for
the stadium \cite{BSS, H, BZ1, BZ2} and for Donnelly's surfaces
\cite{D}. As with eigenfunctions, we can consider the limits
\begin{equation}  \langle
 A
 \psi_j, \psi_j \rangle \to \int_{S^*M} \sigma_A d \nu \end{equation}
of  matrix elements $\langle A \psi_j, \psi_j \rangle$ of
quasimodes.  We say  that the quasi-mode  `scars' if the limit
measure  $d\nu$ has a non-zero singular component  relative to
$dL$.
 For instance, bouncing ball modes of stadia `scar' on the
 Lagrangean manifold with boundary formed by the bouncing ball
 orbits in the central rectangle, and the similar quasi-modes in
 \cite{D} scar on the circles in the cylindrical part. The
 existence of such scarring quasi-modes suggests that these
 systems are not QUE.

To explore this suggestion, we consider  the decomposition of
scarring quasi-modes into sums of true eigenfunctions.

\begin{defin}  \label{ESS}  We say that a quasimode $\{\psi_k \}$ of order $0$ as in (\ref{QM})
with  $||\psi_k||_{L^2} = 1$ has
 $n(k)$ essential frequencies if, for each $k$, there exists a
 subset $\Lambda_k \subset Sp(\Delta) \cap [\mu_k - \delta, \mu_k
 + \delta]$ with  $n(k) = \# \Lambda_k$ and constants $c_{jk} \in
 \C$ such that
\begin{equation} \label{PSIK} \psi_k = \sum_{j: \lambda_j^2 \in \Lambda_k} c_{kj} \phi_j +
\eta_k,\;\; \mbox{with}\;\;  ||\eta_k||_{L^2} = o(1).
\end{equation}
\end{defin}

 The following problems then seem
interesting (the first is implicit in \cite{HO}).

\begin{itemize}

\item  Bound the number $n(k)$ of essential frequencies of a quasimode
$\{\psi_k\}$ of order $0$ which  tends to  a singular (i.e.
non-Liouville) classical limit, e.g. a periodic orbit measure.

\item  Bound the order $s$  of a quasimode with a singular limit
(intuitively,  the diameter of the set of eigenvalues which
composes its packet of eigenfunctions.)

\end{itemize}
In other words, the questions are whether one can build a
quasimode with a singular limit and (i) with anomalously few
essential frequencies, or (ii) with anomalously low order. This
softens the mathematicians' criterion of scarring as the existence
of a sequence of actual modes (eigenfunctions) whose limit measure
$\nu$ in (\ref{QUE}) has a singular component relative to
Liouville measure \cite{S}.

 The following shows that   quasi-modes with a uniformly bounded
 number of essential frequencies and singular limits do not exist
 for QUE systems.

\begin{cor} \label{COR} If there exists a quasi-mode $\{\psi_k\}$ of order $0$
for $\Delta$ as in (\ref{PSIK}) and a constant $C > 0$  with the
properties:
\begin{itemize}

\item  (i)  $n(k)  \leq C,\; \forall \; k$;

\item (ii) $\langle A \psi_k, \psi_k \rangle \to \int_{S^*M} \sigma_A d
\mu$ where $d \mu \not= dL$.

\end{itemize}

Then $\Delta$ is not QUE.
\end{cor}

\begin{proof}

We  argue by contradiction.  If $\Delta$ were QUE, we would have
(in the notation of (\ref{PSIK}):
$$\begin{array}{ll}\langle A \psi_k, \psi_k \rangle & = \sum_{i, j: \lambda_i^2, \lambda_j^2 \in \Lambda_k} c_{kj} \bar{c}_{k i}\langle A
\phi_i, \phi_j \rangle + o(1) \\ & \\&  = \sum_{j: \lambda_j^2 \in
\Lambda_k} |c_{kj}|^2 \langle A \phi_j, \phi_j \rangle +
 \sum_{i \not= j: \lambda_i^2, \lambda_j^2 \in
\Lambda_k} c_{kj} \bar{c}_{k i}\langle A \phi_i, \phi_j
\rangle + o(1)\\ & \\
& = \int_{S^*M} \sigma_A dL + o(1),
\end{array} $$
by Proposition \ref{PONE}. This contradicts (ii). In the last line
we used that $|\lambda_i - \lambda_j| \to 0$ if $ \lambda_i^2,
\lambda_j^2 \in \Lambda_k$ and that  $\sum_{j : \lambda_j^2 \in
\Lambda_k}  |c_{kj}|^2 = 1 + o(1)$, since $||\psi_k||_{L^2} = 1$.

\end{proof}

  The
assumption that $n(k) \leq C$ could be weakened if we knew
something about the rate of decay of the individual elements
$\langle A \phi_j, \phi_k \rangle$ and $|\langle A \phi_j, \phi_j
\rangle - \int_{S^*M} \sigma_A d L|$.

We now consider the implications for the stadium and for
Donnelly's surface. In both cases, it is unknown how many
essential frequencies are needed to build the associated  bouncing
ball quasi-modes. On average,  intervals of fixed width have a
uniformly bounded number of $\Delta$-eigenvalues in dimension $2$,
and this suggests that $n(k) \leq C$. Our result would then show
that such systems are automatically not QUE (as is widely
believed). On the other hand,  the standard remainder estimate for
Weyl's law allows $O(\sqrt{k})$ eigenvalues of $\Delta$  in the
interval $[\mu_k - \delta, \mu_k + \delta]$, and it is possible
that exceptionally high clustering occurs around the bouncing ball
quasi-eigenvalues.  Numerical evidence  \cite{H, HO,  BSS} seems
to show that no exceptional clustering occurs and that bouncing
ball quasi-modes are combinations of a fixed number of actual
modes. But there are at this time no rigorous results of this kind
or in general on the soft scarring criteria above.

We close with some recent references to the literature on
eigenfunction scarring. At this time, no $\Delta$ are  proved to
be QUE and
 none are proved to be non-QUE. But very recently
 E. Lindenstrauss \cite{L} has proved that the special basis of Hecke
 eigenfunctions of arithmetic hyperbolic surfaces has the QUE
 property (see also  \cite{S, RS, BL, W} for prior results in this line). Also,  Faure-Nonnenmacher-de Bievre
 have recently proved that certain quantum cat maps are not QUE \cite{FN, FND}.      Burq-Zworski  \cite{BZ1, BZ2} have
 recently
 given  upper bounds on the concentration of eigenfunctions  in the central
 part of stadia or in collars around  hyperbolic closed geodesics Riemannian manifolds, which show that
 the optimal order of quasimodes with singular concentration in
 these regions is $0$. They further show
 that stadium eigenfunctions cannot scar on
 smaller sets than the entire set of bouncing ball orbits.

\end{document}